\pdfoutput=1

\documentclass[11pt]{article}

\usepackage[final]{acl}

\usepackage{times}
\usepackage{latexsym}
\usepackage{booktabs}
\usepackage{multirow}
\usepackage{amsmath} 
\usepackage[T1]{fontenc}

\usepackage[utf8]{inputenc}

\usepackage{microtype}

\usepackage{inconsolata}

\usepackage{graphicx}

\title{CART: A Generative Cross-Modal Retrieval Framework \\
with Coarse-To-Fine Semantic Modeling}

\author{
 \textbf{Minghui Fang\textsuperscript{1}\footnotemark[1]},
 \textbf{Shengpeng Ji\textsuperscript{1}\footnotemark[1]},
 \textbf{Jialong Zuo\textsuperscript{1}\footnotemark[1]},
 \textbf{Hai Huang\textsuperscript{1}},
 \textbf{Yan Xia\textsuperscript{1}},
 \textbf{Jieming Zhu\textsuperscript{2}\footnotemark[2]}
 \\
 \textbf{Xize Cheng\textsuperscript{1}},
 \textbf{Xiaoda Yang\textsuperscript{1}},
 \textbf{Wenrui Liu\textsuperscript{1}},
 \textbf{Gang Wang\textsuperscript{2}},
 \textbf{Zhenhua Dong\textsuperscript{2}},
 \textbf{Zhou Zhao\textsuperscript{1}\footnotemark[2]}
\\
\\
 \textsuperscript{1}Zhejiang University, \quad
 \textsuperscript{2}Huawei Noah’s Ark Lab
\\
 \href{mailto:minghuifang.zju@zju.edu.cn}{minghuifang.zju@zju.edu.cn} \quad
  \href{mailto:jiemingzhu@ieee.org}{jiemingzhu@ieee.org} \quad
  \href{mailto:zhaozhou@zju.edu.cn}{zhaozhou@zju.edu.cn}
}

\begin{document}
\maketitle

\footnotetext[1]{Equal contribution.}
\footnotetext[2]{Corresponding authors.}

\begin{abstract}
Cross-modal retrieval aims to search for instances, which are semantically related to the query through the interaction of different modal data. Traditional solutions utilize a single-tower or dual-tower framework to explicitly compute the score between queries and candidates, which is challenged by training cost and inference latency with large-scale data. Inspired by the remarkable performance and efficiency of generative models, we propose a generative cross-modal retrieval framework (CART) based on coarse-to-fine semantic modeling, which assigns identifiers to each candidate and treats the generating identifier as the retrieval target. Specifically, we explore an effective coarse-to-fine scheme, combining K-Means and RQ-VAE to discretize multimodal data into token sequences that support autoregressive generation. Further, considering the lack of explicit interaction between queries and candidates, we propose a feature fusion strategy to align their semantics. Extensive experiments demonstrate the effectiveness of the strategies in the CART, achieving excellent results in both retrieval performance and efficiency.
\end{abstract}

\section{Introduction}
In recent years, multimedia data has exploded both in quantity and form, cross-modal retrieval \cite{begin, h1, opt, h2} has become a research hot spot. Cross-modal retrieval tasks, which aim to retrieve relevant data from one modality (e.g. image \cite{clip, openclip}, audio \cite{clap2022, clap2023} or video \cite{clip4clip, cap4video}) based on a query from another modality (e.g. text \cite{crossmodelretrieval}), play a crucial role in various multimedia applications. 

\begin{figure}[]
  \centering
  \includegraphics[width=\linewidth]{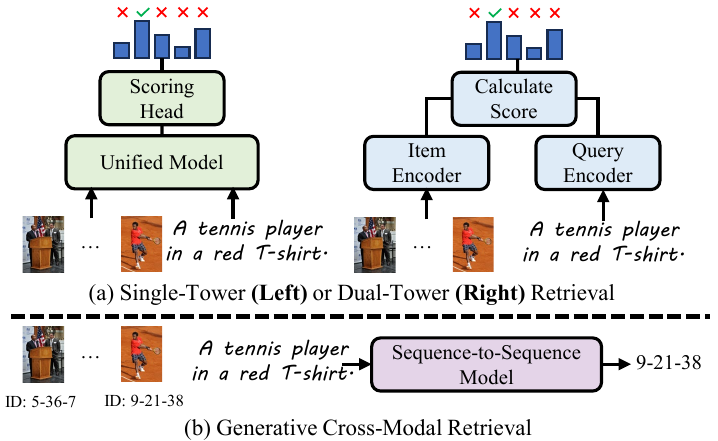}
  \caption{Traditional single-tower / dual-tower retrieval matches the closest candidate to the query by calculating scores, while generative retrieval takes the generating candidate's identifier as the retrieval target.}
  \label{fig:teaser}
\end{figure}

Traditional cross-modal retrieval approaches are mainly based on single-tower or dual-tower framework, as shown in Figure~\ref{fig:teaser} (a). The single-tower model \cite{albef, blip, blip2, internvl} performs fine-grained interactions between queries and candidates in a unified module, which shows excellent capability in retrieval performance, but is difficult to apply to large-scale corpora because of high latency. The dual-tower model \cite{clip, videoclip, clap2023, audioretrieval, imagebind, languagebind} maps different modal data to the joint embedding space via two encoders, which improves the efficiency compared to the former. However, due to the modal gap, the dual-tower model often struggles to align multimodal data \cite{h2} effectively, thus slightly reducing accuracy.

Generative retrieval \cite{DSI, NCI, DSI++, dsi-qg, Tome, multiviewid} is an emerging paradigm in document retrieval that assigns identifiers \cite{Genre, seal, multiviewid} to each candidate, followed by treating the generated identifiers as retrieval targets. This paradigm implements the retrieval process using a sequence-to-sequence model with excellent retrieval accuracy and efficiency, which sheds new light on cross-modal retrieval. As shown in Figure~\ref{fig:teaser} (b), there is no need to maintain a unified embedding space. Leveraging the power of generative models significantly enhances retrieval performance, and the generation speed remains independent of the dataset size.

While such generative framework has shown impressive performance in document retrieval, combining generative retrieval with cross-modal retrieval faces several challenges. Firstly, unlike documents where titles \cite{Ultron, multiviewid}, strings \cite{seal, sedsi}, or keywords \cite{DynamicRetriever, Genre, termasid} can be directly used as identifiers, multimodal data are represented primarily through visual or auditory modalities. This means that we need effective discretization schemes to construct semantic identifiers for multimodal data, which is the core of generative retrieval. Secondly, semantic identifiers constructed from low-level visual or auditory information often exhibit a gap with the high-level semantics of natural language queries. Thirdly, generative retrieval encodes all candidates directly into the model parameters, lacking an explicit interaction process between queries and candidates.

In this work, we propose a generative retrieval framework, denoted as Cross-modal Autoregressive Retrieval Transformer (namely CART), designed to fully support end-to-end text-to-image/audio/video retrieval. The entire framework is divided into three modules: semantic identifier generation, caption enhancement, and feature fusion. Specifically, we propose combining the K-Means and RQ algorithms to generate identifiers with hierarchical semantic information for the candidates. Secondly, we perform caption generation for images/videos/audio, which serves as semantically aligned natural language queries. Finally, we design a two-branch coarse-to-fine feature fusion strategy, which is consistent with the hierarchical structure of semantic identifiers.

We conduct extensive experiments to demonstrate that CART has strong performance in text-to-image/audio/video retrieval. Meanwhile CART has stable retrieval speed on both CPU and GPU, meaning it is not affected by the number of candidates. Ablation studies and in-depth analyses validate the effectiveness and robustness of CART. CART enables end-to-end generative cross-modal retrieval based on a unified differentiable framework, providing a new solution for information retrieval in multimedia applications.

Our contributions are highlighted as follows.
\begin{itemize}
    \item To the best of our knowledge, CART is the first generative cross-modal retrieval framework that comprehensively supports text-to-image/audio/video retrieval.
    \item We propose generating coarse-to-fine semantic identifiers, enabling a unified representation of multimodal data.
    \item We propose a coarse-fine feature fusion mechanism which effectively helps the retrieval model to perceive the query semantics.
    \item Extensive experiments demonstrate the superior performance of CART. Meanwhile, CART has stable retrieval efficiency and beats the single-tower model or dual-tower model under large-scale corpus.
\end{itemize}

\begin{figure*}[t]
  \centering
  \includegraphics[width=0.9\textwidth]{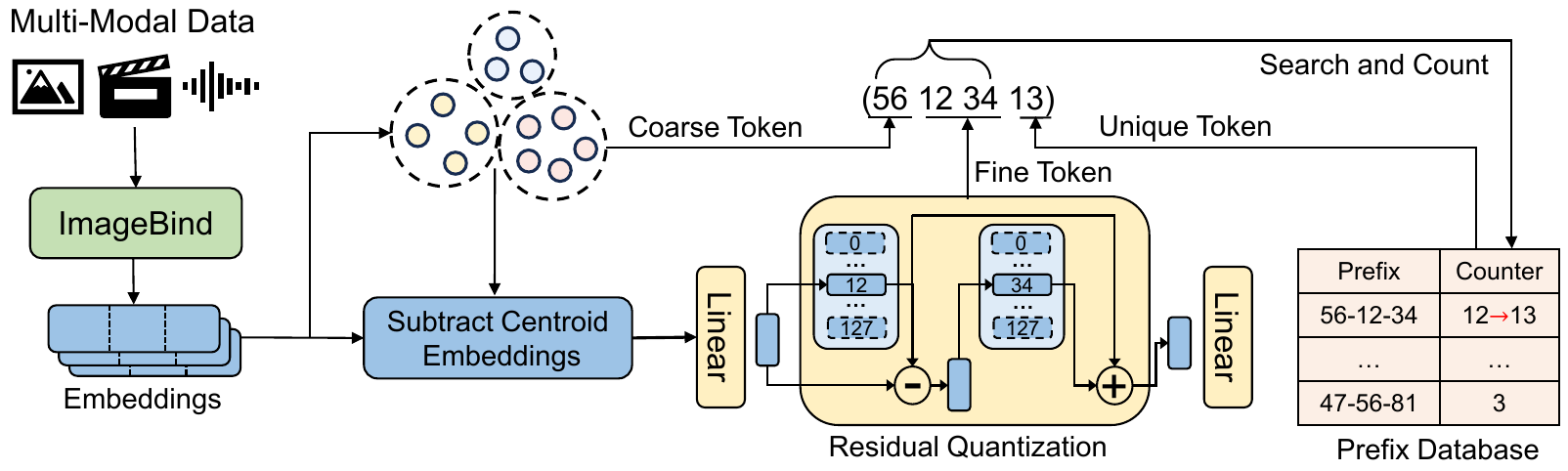}
  \caption{A coarse-Fine semantic identifier generation strategy.}
  \label{fig:gen_id}
\end{figure*}

\section{Related Works}
\subsection{Cross-Modal Retrieval}
Cross-modal retrieval aims to enable searches across different modalities. Currently, the mainstream approaches \cite{clip, openclip, clap2022, clap2023, clip2video,cap4video} to cross-modal retrieval tend to use either single-tower or dual-tower models, both of them formulate cross-modal retrieval as a discriminative problem, relying on discriminative loss and negative samples. 

The difference is that the former performs fine-grained interactions between the query and the candidates within a unified module, e.g., Cross-Attention, after which the score of each candidate is obtained by mapping in the output layer. BLIP \cite{blip}, BLIP-2 \cite{blip2}, InternVL-G \cite{internvl} follow this setting to achieve high-accuracy retrieval in the domain of text-to-image retrieval; whereas, the latter tends to utilize two independent encoders to map different modalities into a joint feature space and then calculate their similarity by a distance function. CLIP \cite{clip}, CLAP \cite{clap2022} and CLIP2Video \cite{clip2video} follow this setting and complete the pre-training on large-scale datasets, achieving remarkable outcomes in text-image retrieval \cite{text2image}, text-audio retrieval \cite{text2audio} and text-video retrieval \cite{text2video}, respectively.

Both solutions have their own trade-offs; the single-tower model can achieve extremely high accuracy, but sacrifices efficiency because of fine-grained interactions, leading it to be suitable only for small-scale retrieval tasks, and the dual-tower model effectively improves speed, but demands extensive data to train a unified representation space, leading to a slight decrease in retrieval accuracy. In contrast, CART aims to utilize generative models to improve retrieval efficiency while ensuring excellent retrieval performance.

\subsection{Generative Retrieval}
Generative retrieval \cite{DSI, NCI, multiquery} represents an innovative paradigm in document retrieval. Its core idea is to construct indices for documents and replace the process of calculating relevance scores with generating identifiers.

The performance of generative retrieval is heavily influenced by candidate identifiers. Past studies have delved into the application of various identifier types across diverse scenarios, including keywords-based \cite{Ultron, DynamicRetriever, Genre, termasid, MIRA}, Web URLs \cite{Tome, llmforsearch}, and substrings of paragraphs \cite{Ultron, seal, sedsi, multiviewid}. Considering the absence of explicit document content, DSI \cite{DSI} and NCI \cite{NCI} recognize the challenge in generating document identifiers solely based on input queries. Therefore, they advocate using the hierarchical K-Means algorithm to inject prior knowledge into identifiers. In other words, documents sharing close semantic ties are assigned similar docids. This methodology seamlessly embeds the semantic information of documents into the decoding process, which facilitates the learning process of the retrieval model. CART considers to build a unified identifier generation pipeline for multimodal data, combining K-Means and residual vector quantization to compress the embedding within a fixed step size. In addition, we investigate the application of generative retrieval techniques to recommendation tasks~\cite{EAGER,EAGER-LLM}.

\section{Method}
\subsection{Overview}
Overall, CART is a framework that comprehensively supports generative text-to-image/video/audio retrieval, based on a coarse-to-fine semantic modeling scheme. The whole pipeline can be divided into three modules. (1) \textbf{Semantic Identifier Construction}: constructing semantically rich \textbf{identifiers} for all candidates based on the discretization method; (2) \textbf{Caption Enhancement}: generating captions for multimodal data as \textbf{queries}; (3) \textbf{Feature Fusion}: generative models are trained using the \textbf{<query, identifier> pairs}, and with the help of effective feature fusion mechanism to improve the performance. To enhance the robustness of CART, we incorporate consistency loss to suppress overfitting, and guarantee to generate valid identifiers via constrained beam search.

\subsection{Coarse-Fine Identifier Generation}
Images, audio, and video lack explicit attributes that can be directly used as identifiers, necessitating their discretization into token sequences in the latent space. The pipeline is illustrated in Figure~\ref{fig:gen_id}.

\subsubsection{Coarse Token} Intuitively, the first token of identifier is critical, if the first token of identifier is generated incorrectly, subsequent generation will be meaningless. With this in mind, we hope that the first token of the identifier captures the full semantic information of the item, so that the retrieval model can easily predict the semantically closest information based on a natural language query. We utilize K-Means \cite{kmeans} to cluster the embeddings of all the items, which are encoded from ImageBind \cite{imagebind}, and the clustering result is considered as the first token of the semantic identifier. Although the K-Means algorithm can finish item categorization rapidly in the whole data space, it is difficult to consider the subtle gaps between items effectively. Therefore, we denote the clustering result $k$ of K-Means as the coarse token.

\subsubsection{Fine Token} Subsequently, we utilize original embeddings to subtract the embedding from the clustering center of the K-Means algorithm, in order to highlight the subtle differences between items. Then we use RQ-VAE \cite{soundstream} to construct the remaining tokens in the semantic identifier. 

RQ-VAE combines the advantages of variational autoencoder and residual quantization, where the autoencoder is jointly trained by updating the DNN encoder-decoder parameters and the quantization codebook. The encoder $E(\cdot)$ first downsamples the input $x$ to learn the latent representation $z = E(x)$, which removes noise and unimportant information and preserves the most meaningful features of the data. The quantizer $Q = \{C_1, C_2, \ldots, C_M\}$ where $M$ is the number of codebooks. The codebook $C_m = \{e_m^1, e_m^2, \ldots, e_m^N\}$ where $N$ is the codebook size and $e_m^n$ denotes the $n$-th entry in the $m$-th codebook. The initial residual of the quantizer is defined as $r_0 = z$, then $r_0$ is quantized by mapping it to the nearest embedding $e_1^k$ from the first codebook ${C_1}$. The index of the closest embedding, i.e., $v_1 = argmin_k \Vert{r_0 - e_1^k}\Vert$, represents the token of the semantic identifier obtained in codebook $C_1$, the residual is updated to $r_1 = {r_0 - e_1^{v_1}}$. This process is executed recursively $m$ times and we obtain the index list $V = \{v_1, v_2, \ldots, v_M\}$ denoting the semantic identifier produced by the RQ-VAE. Note that all codebooks do not share parameters, they are updated independently to distinguish between the different semantics of each hierarchy. Finally $Q(z) = \sum_{i=1}^Me_i^{v_i}$ is passed to the decoder $D(\cdot)$ to try to reconstruct the input $x$. We train the RQ-VAE using two loss functions $\mathcal{L} = \mathcal{L}_{recon} + \alpha \mathcal{L}_{commit}$, where: 
\begin{equation}
\begin{split}
    \mathcal{L}_{recon} &= \Vert x-D(Q(E(x))) \Vert^2_2, \\
    \mathcal{L}_{commit} &= \sum_{i=1}^{M} \Vert sg[r_i] - e_{v_i}^{i} \Vert^2_2 + \beta \Vert r_i - sg[e_{v_i}^{i}] \Vert^2_2,
\end{split}
\end{equation}
Here, both $\alpha$ and $\beta$ are hyperparameters, and $sg$ is the stop-gradient operation \cite{sg}. RQ-VAE hierarchically computes the residuals of the features for each item. In the refinement process, it focuses on the variation of the features at different levels and can capture the subtle features of the items. Therefore, we denote the quantization result of RQ-VAE as the fine token.

\subsubsection{Unique Token} In large-scale corpora, it is common to have semantically close items, which makes it difficult to avoid identifier collisions \cite{recrvq}. In order to assign unique identifiers to each item, we maintain a prefix database to detect conflicts. We perform a "search and count" operation on all identifiers, if the identifier already exists, increase its counter by one, on the contrary, insert the identifier into the prefix database with an initial counter value of 0. The value of the counter $u$ will be appended to the end of the identifier as an additional token to ensure that each identifier is unique. Note that search, insert or update of the prefix database is so efficient that its time spent is even negligible.

Finally, we concatenate coarse token, fine token and unique token as identifiers for items, i.e. $(k, v_1, v_2, \cdots, v_M, u)$.

\subsection{Caption Enhancement}
One huge challenge of generative retrieval is how to make the retrieval model aware of the semantic information represented by the identifiers. Since the content of each item is not explicitly known at inference, it must be incorporated into the model parameters during training, as indicated in previous work \cite{NCI, dsi-qg, mevi} . 

To bridge this gap, we employ pre-trained multimodal models to generate captions for each candidate, taking information-rich captions as queries. The retrieval model is subsequently trained using <query, identifier> pairs, which effectively understands the information embedded in each token in the semantic identifier.

\begin{figure*}[t]
  \centering
  \includegraphics[width=0.9\textwidth]{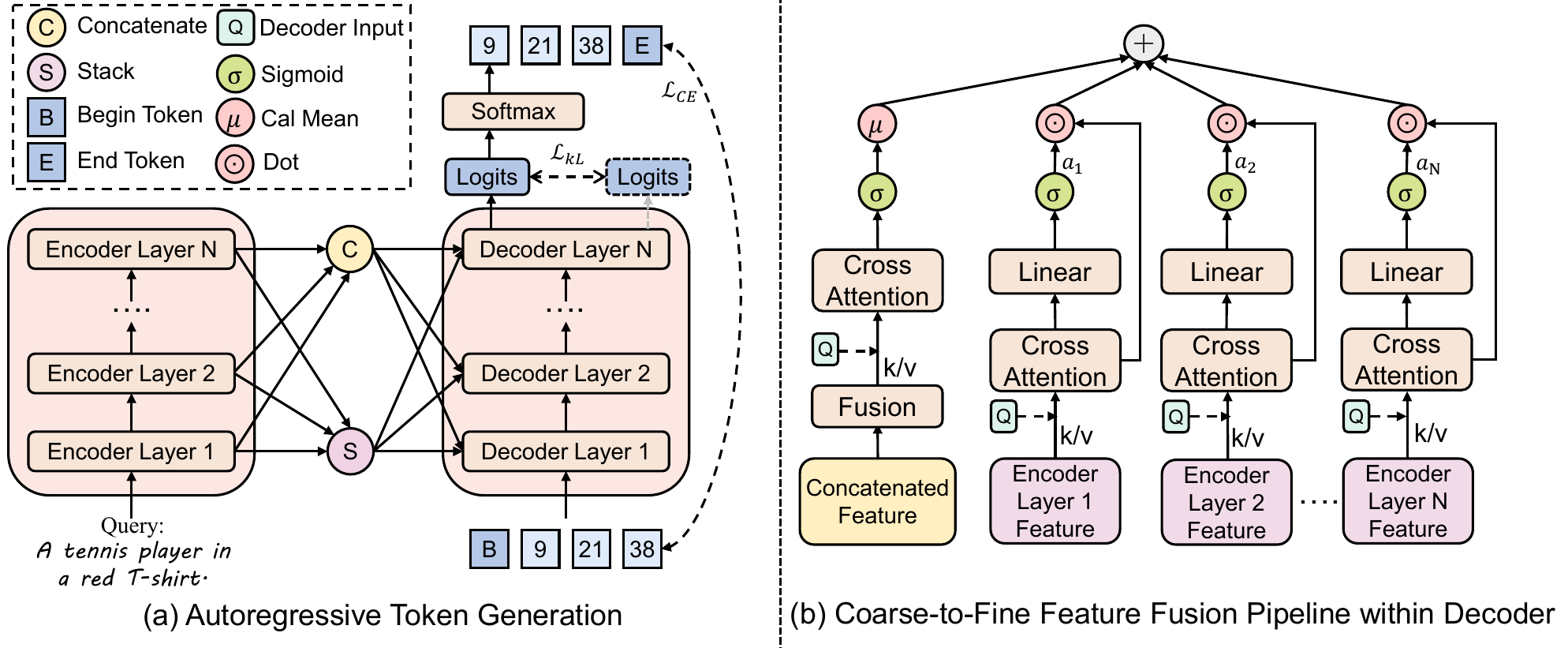}
  \caption{The architecture of the CART.}
  \label{fig:pipeline}
\end{figure*}

\subsection{Coarse-to-Fine Feature Fusion}
CART utilizes a standard encoder-decoder architecture, both made of stacks of attention layers. The encoder is in charge of capturing the semantics of the input query, the decoder reads the output of the encoder and generates semantic identifier token by token. Given an input query $q$, the output of an encoder with $S$ layers can be defined as $E(q) = (E_1, E_2, \ldots, E_S)$. Each encoder layer can capture different semantic representations \cite{meshedtransformer} of queries. The lower layer can capture some basic regional relevance, while the higher layer further refines and optimizes semantic representations based on the former. In light of this, we propose the coarse-to-fine feature fusion strategy. As shown in Figure~\ref{fig:pipeline}, we process the encoder output in two branches.

\subsubsection{Coarse Fusion} The outputs from each encoder layer are concatenated and passed through a fusion layer, to integrate information from different levels of the encoding process. Because of the "concatenate-then-fusion" method, we naturally call it coarse fusion, which is similar to the idea of coarse token. Formally,
\begin{equation}
    Z = W[E_1, E_2, \ldots, E_S]+b,
\end{equation}
where $[\cdot, \cdot]$ denotes connecting the outputs of all encoder layers from the last dimension, $W$ and $b$ are the learnable weights and bias, respectively. Subsequently, $Z$ interacts with the decoder input $Y$ through cross-attention and completes the post-processing using self-gating. The whole process of coarse fusion $G(\cdot, \cdot)$ can be written as
\begin{equation}
\begin{split}
    C(Y, Z) &= \mathrm{Attention}(W_qY, W_kZ, W_vZ), \\
    G(Y, Z) &= \frac{1}{S}\sigma(C(Y, Z)),
\end{split}
\end{equation}
where $\mathcal{C}(\cdot, \cdot)$ stands for the encoder-decoder cross-attention, $S$ is the number of encoder layers and $\sigma(\cdot)$ is the sigmoid activation. $W_q$, $W_k$ and $W_v$ are matrices of learnable weights.

\subsubsection{Fine Fusion} The purpose of fine fusion is for more careful deciding which subtle features to use. Intuitively, similar to mixture-of-experts \cite{moe}, this design treats each encoder layer as an expert. The outputs of each layer in the encoder interact independently with the inputs of the decoder through cross-attention, while utilizing the weights $\alpha_i$ regulate each encoder layer's contribution and their relative importance. The decision-making process $L(\cdot, \cdot)$ can be written as
\begin{equation}
    L(Y, E(q)) = \sum_{i=1}^S \alpha_i \odot \mathcal{C}(Y, E_i),
\end{equation}
and inspired by \cite{meshedtransformer}, we introduce the memory-augmented mechanism for weight computation, as follows
\begin{equation}
    \alpha_i = \sigma(W_i[Y, \mathcal{C}(Y, E_i)] + b), 
\end{equation}
where $[\cdot, \cdot]$ indicates concatenation, $\sigma(\cdot)$ is the sigmoid activation, $W_i$ is the learnable weight matrices and $b$ is bias. 

Finally, we add the coarse feature and fine feature as input to the next decoder layer.

\subsection{Training and Inference}
We use <query, identifier> pairs to train a sequence-to-sequence model. Given an input query $q$, the probability of generating the semantic identifier $T$ can be written:
\begin{equation}
    p(T \mid E(q), \theta) = \prod_{i=1}^Jp(t_i \mid E(q), t_{<i}, \theta_i),
\end{equation}
where $t_i$ is the $i$-th token in the semantic identifier $T$, $J$ denotes the length of $T$ and $E(\cdot)$ denotes the encoder, $\theta$ denotes the total parameters and $\theta_i$ is the parameter for the $i$-th step.

To mitigate the overfitting problem, we introduce a bidirectional KL divergence loss \cite{r-drop} to train the decoding process. For the user query, we denote the decoder representations by two forward passes with independent dropouts before softmax as $P$ and $Q$, i.e., 
\begin{equation}
    \mathcal{L}_{KL} = \sum_{i} P(i) \log \left( \frac{P(i)}{Q(i)} \right) + \sum_{i} Q(i) \log \left( \frac{Q(i)}{P(i)} \right),
\end{equation}

As shown in Figure~\ref{fig:pipeline}, given a collection of training examples $\mathcal{D} = \{(q, d)\}$ composed of queries (training queries and augmented queries) and identiﬁers, the loss function can be written as follows:
\begin{equation}
    \mathcal{L}(\theta) = \sum_{(q,d)\in\mathcal{D}}(log p(d \mid E(q), \theta) + \omega \mathcal{L}_{KL}),
\end{equation}
where $p(d \mid E(q), \theta)$ denotes the probability of generating $d$ with $q$ as the input and $\omega$ denotes a scaling factor of KL divergence loss.

In the inference stage, we execute a constrained beam search on the decoder network. Benefiting from the prefix database constructed when generating identifiers, we can effectively build a prefix tree for all identifiers, which will restrict the model to generating only valid identifiers. Due to the hierarchical nature of identifiers, it is convincing \cite{NCI} to constrain the beam search decoding process with a prefix tree.

\section{Experiments}
We evaluate the performance of CART on text-to-image/audio/video, respectively. 

\subsection{Datasets \& Baselines}
\paragraph{Datasets} 
We selected six widely used datasets: Flickr30K  \cite{flickr}, MS-COCO \cite{mscoco}, Clotho \cite{clotho}, AudioCaps \cite{audiocaps}, MSR-VTT \cite{msrvtt} and MSVD \cite{msvd}. Further details are provided in Appendix~\ref{sec:datasets}.

\paragraph{Baselines} We selected single-tower, dual-tower, and generative retrieval architectures as baselines. Further details are provided in Appendix~\ref{sec:baseline}.

\subsection{Implementation Details}
The implementation mainly consists of the semantic identifier generation pipeline and the retrieval pipeline, with details on network structure and parameter settings provided in Appendix~\ref{sec:sig}. All experiments are based on 4 NVIDIA V100 GPUs with training batch size set to 256.

\subsection{Empirical Results}
\subsubsection{Metrics} Consistent with prior studies, we use widely accepted metrics for information retrieval, including Recall$@K$ and Mean Reciprocal Rank (MRR). The details about the metrics are in Appendix~\ref{sec:metrics}.

\begin{table}[!t]
\centering
\small
\caption{Compared to single-tower models, results from original papers.}
\resizebox{\columnwidth}{!}{
\begin{tabular}{lccc|ccc|c}
\toprule
\multirow{2}{*}{Method} & \multicolumn{3}{c|}{Flickr30K}               & \multicolumn{3}{c|}{MS-COCO}                & \multirow{2}{*}{Throughput} \\ \cline{2-7}
                        & R@1            & R@5           & R@10           & R@1            & R@5           & R@10           &                           \\
\midrule
BLIP-2                 & 89.7           & 98.1          & 98.9           & 66.3           & 86.5          & 91.8           & 1.68/s                    \\
InternVL-G             & 85.0   & 97.0  & 98.6   & 58.6   & 81.3  & 88.0   & 2.03/s                    \\
CART(Ours)              & 81.8   & 96.1  & 98.4   & 52.4  & 77.5 & 86.1  & 105.8/s                   \\
\bottomrule
\end{tabular}}
\label{tab:compare_with_single}
\end{table}

\subsubsection{Comparison of single-tower retrieval} As shown in Table~\ref{tab:compare_with_single}, We observe that the single-tower model has an overwhelming advantage on R@1, thanks to the fine-grained interactions between queries and candidates. Notably, with the increase in the number of recalls, the gap between CART and BLIP-2 \cite{blip2} narrows faster than that between InternVL-G \cite{internvl} and BLIP-2, which proves that CART has comparable performance. Afterwards, we simulated 1M candidate images and experimented with 100 concurrent queries. Table~\ref{tab:compare_with_single} shows the huge latency of the single-tower model, which is not applicable to large-scale retrieval. In contrast, CART achieves a better balance between performance and efficiency.

\begin{table*}[ht]
\centering
\small
  \caption{Compared to dual-tower models, the results are from the original papers(*) or official reproduction.}
  \label{tab:results}
  \resizebox{\textwidth}{!}{
  \begin{tabular}{ccccccccccc}
    \toprule
    \multirow{2}{*}{Task} & \multirow{2}{*}{Method} & \multicolumn{4}{c}{Flickr30k} && \multicolumn{4}{c}{MS-COCO} \\ \cline{3-6} \cline{8-11} 
    && R@1 & R@5 & R@10 & MRR@10 && R@1 & R@5 & R@10 & MRR@10 \\
    \midrule
    \multicolumn{1}{c}{\multirow{7}{*}{Text-to-Image Retrieval}} 
    & CLIP & 55.9 & 82.8 & 90.6 & 67.3 && 30.4 & 55.9 & 66.8 & 41.3 \cr
    & OpenCLIP & 63.9 & 87.3 & 93.2 & 74.1 && 39.4 & 65.4 & 75.6 & 50.4 \cr
    & MobileCLIP & 74.9 & 92.9 & 96.3 & 82.6 && 50.3 & 74.9 & 82.4 & 60.8 \cr
    & ONE-PEACE(Pretrained) & 73.4 & 91.5 & 95.4 & 81.2 && 48.0 & 71.6 & 79.5 & 58.0 \cr
    & ImageBind(Huge) & 74.9 & 93.0 & 96.1 & 82.7 && 49.4 & 73.3 & 81.5 & 59.6 \cr
    & LanguageBind(Image) & 69.5 & 90.8 & 94.9 & 78.7 && 45.3 & 69.8 & 78.6 & 55.8 \cr
    & CART(ours) & \textbf{81.8} & \textbf{96.1} & \textbf{98.4} & \textbf{88.0} && \textbf{52.4} & \textbf{77.5} & \textbf{86.1} & \textbf{63.2} \\
    \midrule
    \multirow{2}{*}{Task} & \multirow{2}{*}{Method} & \multicolumn{4}{c}{Clotho} && \multicolumn{4}{c}{AudioCaps} \\ \cline{3-6} \cline{8-11}  
    && R@1 & R@5 & R@10 & MRR@10 && R@1 & R@5 & R@10 & MRR@10 \\
    \midrule
    \multicolumn{1}{c}{\multirow{5}{*}{Text-to-Audio Retrieval}}
    & CLAP-HTSAT* & 16.1 & 38.3 & 51.1 & - && 36.1 & 71.8 & 83.9 & - \cr
    & HTSAT-BERT* & 19.7 & 45.7 & 59.4 & - && 42.2 & 76.5 & 87.1 & - \cr 
    & VALOR-B* & 17.5 & 42.7 & 55.3 & - && 40.1 & 73.9 & 83.1 & - \cr
    & ONE-PEACE(Pretrained)* & 22.4 & 49.0 & 62.7 & - && 42.5 & 77.5 & \textbf{88.4} & - \cr 
    & LanguageBind(Audio) & 16.3 & 40.4 & 53.9 & 26.6 && 15.2 & 49.8 & 67.6 & 29.8 \cr
    & CART(ours) & \textbf{46.4} & \textbf{70.6} & \textbf{76.0} & \textbf{52.3} && \textbf{49.8} & \textbf{79.8} & 86.8 & \textbf{62.0} \\
    \midrule
    \multirow{2}{*}{Task} & \multirow{2}{*}{Method} & \multicolumn{4}{c}{MSR-VTT} && \multicolumn{4}{c}{MSVD} \\ \cline{3-6} \cline{8-11}  
    && R@1 & R@5 & R@10 & MRR@10 && R@1 & R@5 & R@10 & MRR@10 \\
    \midrule 
    \multicolumn{1}{c}{\multirow{7}{*}{Text-to-Video Retrieval}}
    & CLIP2Video* & 45.6 & 72.5 & 81.7 & - && 47.0 & 76.8 & 85.9 & -  \cr
    & CLIP4Clip* & 44.5 & 71.4 & 81.6 & - && 46.2 & 76.1 & 84.6 & -  \cr
    & UMT-B* & 46.3 & 72.7 & 82.0 & - && 47.4 & 76.8 & 84.0 & - \\
    & Cap4Video* & 49.3 & 74.3 & 83.8 & - && 51.8 & 80.8 & 88.3 & -  \cr
    & ImageBind(Huge) & 36.8 & 61.8 & 70.0 & 46.4 && 39.3 & 67.5 & 78.5 & 53.4  \cr
    & LanguageBind(Video) & 42.7 & 67.1 & 77.0 & 54.4 && 53.5 & 80.5 & 87.5 & 60.6  \cr
    & CART(ours) & \textbf{52.6} & \textbf{75.4} & \textbf{84.2} & \textbf{61.8} && \textbf{63.6} & \textbf{87.9} & \textbf{92.8} & \textbf{73.6} \\
    \bottomrule
  \end{tabular}}
\end{table*}

\begin{table}[!ht]
\centering
\caption{Compared to generative retrieval models.}
\resizebox{\columnwidth}{!}{
\begin{tabular}{ccccclccc}
\toprule
\multirow{2}{*}{Paradigm} & \multirow{2}{*}{Method} & \multicolumn{3}{c}{Flickr30K} &  & \multicolumn{3}{c}{MS-COCO} \\ \cline{3-5} \cline{7-9} 
                          &                          & R@1      & R@5      & R@10    &  & R@1     & R@5     & R@10    \\
\midrule
\multirow{3}{*}{GRACE}    & Semantic ID              & 22.9     & 34.9     & 37.4    &  & 13.3    & 30.4    & 35.9    \\
                          & Structured ID            & 37.4     & 59.5     & 66.2    &  & 16.7    & 39.2    & 50.3    \\
                          & Atomic ID                & 68.4     & 88.9     & 93.7    &  & 41.5    & 69.1    & 79.1    \\
\midrule
CART(Ours)                       & Coarse-Fine ID           & \textbf{81.78}    & \textbf{96.14}    & \textbf{98.38}   &  & \textbf{52.42}   & \textbf{77.53}   & \textbf{86.11}  \\
\bottomrule
\end{tabular}}
\label{tab:compare_grace}
\end{table}

\subsubsection{Comparison of dual-tower retrieval}
As shown in in Table~\ref{tab:results}, we report the retrieval results for CART and the corresponding baselines. In text-to-image/video/audio retrieval, CART exhibits strong performance, leading the baseline in all metrics. Compared to images, videos and audios are temporal and thus contain more semantic information, which increases the difficulty of retrieval. While CART achieves promising results benefiting from coarse-fine semantic identifiers and effective training strategies, proving the feasibility of generative retrieval. We also performed a more detailed efficiency comparison with the dual-tower model, as shown in Figure~\ref{fig:latency}.

\subsubsection{Comparison of generative retrieval} As shown in Table~\ref{tab:compare_grace}, GRACE \cite{dis} performs relatively poorly because the multiple identifier schemes it employs are predefined and do not comprehensively consider semantic information, whereas CART employs a novel coarse-to-fine-grained identifier generation pipeline to effectively facilitate model learning. It is worth noting that GRACE adopts Flamingo \cite{alayrac2022flamingo} (a MLLM) as the model backbone, while CART adopts a naïve small-parameter transformer, the final difference in retrieval results is enough to prove the importance of identifier for generative retrieval.

\subsubsection{Ablation experiment} Furthermore, to investigate the effect of each component, we report the ablation results in Table~\ref{tab:ablation}. The detailed analysis is described below.

\paragraph{w/o consistency loss} This setting removes the consistency-based regularization loss, and we observe that the model is more prone to overfitting, which leads to degradation of performance.

\paragraph{w/o fusion strategy} This setting removes the process of feature fusion in the decoder, demonstrating the importance of fully considering outputs from all encoder layers.

\paragraph{w/o K-Means} This setting removes the process of K-Means and uses only RQ-VAE to construct semantic identifiers. Experiments show that clustering the original embeddings via K-Means brings prior knowledge that is effective for retrieval.

\paragraph{w/o RQ-VAE} This setting removes the process of RQ-VAE and uses only the hierarchical K-Means to construct K-fork trees for the original embeddings, where the paths from the root node to the leaf nodes are considered as identifiers of the items. Experiments show that hierarchical K-Means loses semantic information between different clusters, which is consistent with \citet{recrvq}.

\begin{table}[t]
\centering
\small
  \caption{Ablation Study on Flickr30k, we leave the other ablation results in Appendix~\ref{sec:ablation}.}
  \label{tab:ablation}
  \resizebox{\columnwidth}{!}{
  \begin{tabular}{ccccc}
    \toprule
    Setting & R@1 & R@5 & R@10 & MRR@10 \\
    \midrule
    w/o consistency loss & 81.64 & 95.94 & 98.04 & 87.85 \\
    w/o fusion strategy & 75.54 & 93.54 & 96.72 & 83.11 \\
    w/o K-Means & 79.50 & 94.92 & 97.52 & 86.12 \\
    w/o RQ-VAE & 76.22 & 92.86 & 96.16 & 83.31 \\
    CART & \textbf{81.78} & \textbf{96.14} & \textbf{98.38} & \textbf{88.04} \\
    \bottomrule
  \end{tabular}}
\end{table}

\begin{figure}[t]
  \centering
  \includegraphics[width=0.8\linewidth]{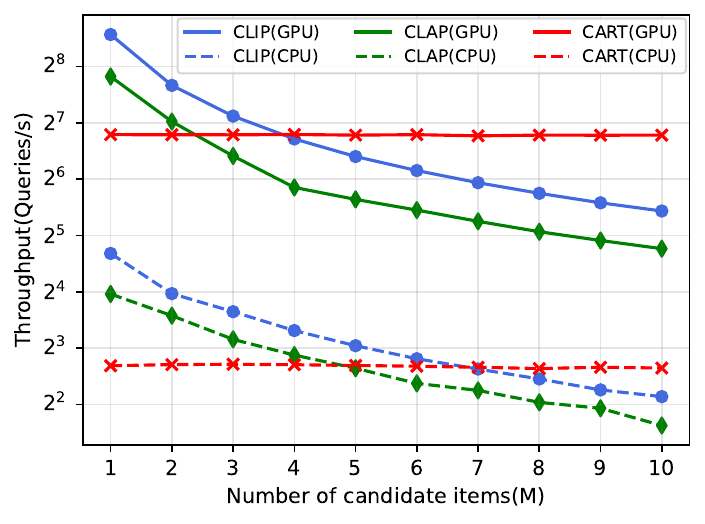}
  \caption{The efficiency of CLIP, CLAP and CART are measured by throughput (queries processed per second).}
  \label{fig:latency}
\end{figure}

\subsection{In-depth Analysis}
\paragraph{Efﬁciency Analysis} We evaluate the retrieval efficiency of CLIP \cite{clip}, CLAP \cite{clap2023}, and CART on both a CPU and an NVIDIA V100 GPU, simulating 100 concurrent queries while varying the number of candidates. The metric is throughput (Queries/per second) and the detailed results are shown in Figure~\ref{fig:latency}. CLIP and CLAP pre-encode candidates into embeddings, where the primary computational costs arise from text encoding, similarity computation, and ranking. In contrast, CART employs beam search to generate identifiers, integrating retrieval and ranking. As the number of candidates increases, the efficiency of dual-tower models like CLIP and CLAP declines due to increased computational demands. However, CART maintains stable efficiency on both GPU and CPU, with its advantage becoming more pronounced as candidate numbers grow, attributed to encoding all items directly as parameters.

\begin{figure}[t]
\centering
  \includegraphics[width=\linewidth]{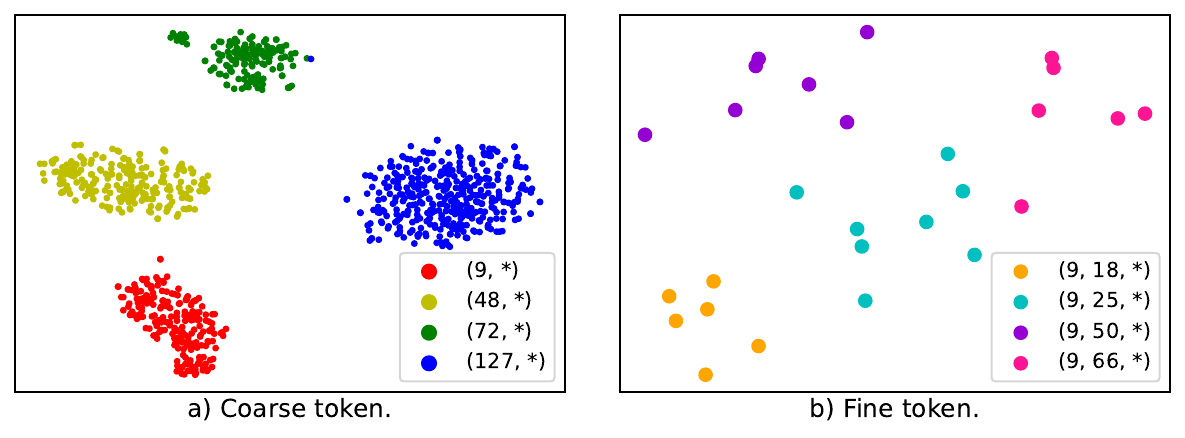}
  \caption{The t-SNE visualization of item embeddings which have the same token prefixes.}
  \label{fig:distribution}
\end{figure}

\begin{table}[t]
  \caption{Results with varying encoder layers.}
  \centering
  \small
  \label{tab:layers}
  \begin{tabular}{c|c|c|c|c}
    \toprule
    Layers & R@1 & R@5 & R@10 & MRR@10 \\
    \midrule
    1 &  76.16 & 94.48 & 97.78 & 84.11 \\
    2 &  80.52 & 95.16 & 97.92 & 86.87 \\
    3 &  81.78 & \textbf{96.14} & \textbf{98.38} & \textbf{88.04} \\
    4 &  \textbf{82.10} & 95.68 & 98.22 & 88.11\\
    5 &  81.08 & 95.64 & 97.78 & 87.37 \\
    \bottomrule
  \end{tabular}
\end{table}

\begin{table}[t]
\centering
\small
\caption{Results with different identifier sizes.}
\label{tab:size}
\begin{tabular}{c|c|ccc|c}
\toprule
K & Codebook & R@1 & R@5 & R@10 & MRR@10 \\
\midrule
\multirow{3}{*}{64} & 64*64 & 75.76 & 92.91 & 96.48 & 83.12 \cr
& 128*128 & 81.22 & 95.52 & 97.54 & 87.39 \cr
& 256*256 & 79.84 & 95.58 & 98.18 & 86.44 \\
\midrule
\multirow{3}{*}{128} & 64*64 & 78.64 & 93.96 & 97.08 & 85.31\cr
& 128*128 & \textbf{81.78} & \textbf{96.14} & \textbf{98.38} & \textbf{88.04} \cr
& 256*256 & 80.86 & 95.61 & 98.01 & 87.26 \\
\midrule
\multirow{3}{*}{256} & 64*64 & 77.68 & 94.81 & 97.62 & 84.93 \cr
& 128*128 & 81.54 & 95.81 & 98.31 & 87.68 \cr
& 256*256 & 80.48 & 95.96 & 98.34 & 87.21\\
\bottomrule
\end{tabular}
\end{table}

\paragraph{Feature Fusion Configuration} 
The effectiveness of coarse-to-fine feature fusion is influenced by the number of encoder layers. We evaluate configurations with \{1,2,3,4,5\} layers, with results summarized in Table~\ref{tab:layers}. The metrics show consistent performance improvement as the number of layers increases, validating the fusion approach. However, excessive layers may lead to overfitting and performance degradation. Consequently, CART employs a 3-layer encoder for optimal balance. Correspondingly, we provide an analysis of the decoder layer in Appendix~\ref{sec:decoder}.

\begin{table}[!ht]
\centering
\small
\caption{Results with different identifier lengths.}
\label{tab:length}
\begin{tabular}{c|c|c|c|c}
\toprule
Length & R@1 & R@5 & R@10 & MRR@10 \\
\midrule
4 & 81.78 & 96.14 & 98.38 & 88.04\\
5 & \textbf{81.84} & \textbf{96.38} & \textbf{98.44} & \textbf{88.51} \\
6 & 80.54 & 95.44 & 98.06 & 87.57 \\
\bottomrule
\end{tabular}
\end{table}

\paragraph{Distribution of Semantic Identifiers} We analyzed the distribution of candidate embeddings with the same token prefix. Figure~\ref{fig:distribution} shows tighter intra-cluster distributions and more dispersed inter-cluster distributions, indicating that the identifiers capture the semantics of the candidates.

\paragraph{Identifier Configuration} We investigate the impact of semantic identifier length and size through experiments on Flickr30k. Results in Tables~\ref{tab:size} and \ref{tab:length} highlight the robustness of CART, with further analysis detailed in Appendix~\ref{sec:idconfig}. We also provide an analysis of the codebook vector dimensions in Appendix~\ref{sec:dim}.

\section{Conclusion}
In this work, we propose CART, a novel generative cross-modal retrieval framework that can fully support text-to-image/video/audio retrieval. Benefiting from the design of coarse-fine semantic identifiers and feature fusion strategy, CART can effectively understand user queries and accurately retrieve candidates. We conducted extensive experiments to demonstrate the effectiveness and robustness of the framework, along with in-depth analysis for a comprehensive understanding of CART. Compared to single-tower and dual-tower frameworks, CART achieves a fine balance between retrieval performance and efficiency. We sincerely hope this will inspire future work.

\section*{Limitation}
We initially explored the potential of generative models to accomplish cross-modal retrieval. However, in reality corpora are usually dynamically changing , and we would like to introduce continuous learning to accommodate this setting in our next work. Meanwhile, the current version only supports natural language queries, and we would like to explore a unified embedding space to accommodate multimodal data queries.

\section*{Acknowledgement}
We thank MindSpore (\url{https://www.mindspore.cn}) for the partial support of this work, which is a new deep learning computing framework.

\bibliography{acl_latex}

\appendix

\section{Datasets}
\label{sec:datasets}

\paragraph{Text to Image Retrieval} Flickr30K \cite{flickr} contains 31,783 images and each image is associated with 5 human-annotated sentences. We select 29,783 images for training, 1000 images for validation and 1000 images for testing. MS-COCO \cite{mscoco} comprises 123,287 images, and each image comes with 5 sentences of annotations. We followed the data split proposed in \cite{mscocosplit} and utilized 113,287 images for training, 5000 images for validation and 5000 images for testing.

\paragraph{Text to Audio Retrieval} Each audio in Clotho \cite{clotho} has 5 manually annotated captions, and there are 3839 audios for training, 1045 audios for validation, and 1045 audios for testing. Since some of the videos in YouTube are no longer available, we do not have the full collection of AudioCaps \cite{audiocaps}. Finally, we have 37,869 audios for training, 384 audios for validation, and 737 audios for testing.

\paragraph{Text to Video Retrieval} MSR-VTT \cite{msrvtt} is a dataset composed of 10,000 videos, each with a length that ranges from 10 to 32 seconds and 200,000 captions. We use 9,000 videos for training that follows the data split of \cite{train9k}, and the test set consists of 1,000 video-text pairs from \cite{test1k}. MSVD \cite{msvd} contains 1,970 videos, each with a length that ranges from one to 62 seconds. Train, validation and test splits contain 1,200, 100, and 670 videos, respectively.
This is an appendix.

\section{Baselines}
\label{sec:baseline}
The single-tower model has powerful performance but high latency, which is not suitable for large-scale retrieval, so we only select BLIP-2 \cite{blip2} and InternVL-G \cite{internvl} as baseline. We focus on comparing the dual-tower models, including the classic CLIP \cite{clip}, as well as the iteratively upgraded OpenCLIP \cite{openclip}, MobileCLIP \cite{mobileclip}, CLAP \cite{clap2023, HTSAT-BERT}, VALOR-B \cite{valor}, CLIP2Video \cite{clip2video}, CLIP4Clip \cite{clip4clip}, UMT-B \cite{umt}, Cap4Video \cite{cap4video}, ImageBind \cite{imagebind}, LanguageBind \cite{languagebind} and ONE-PEACE \cite{onepeace}, which are utilized to perform text-to-picture/video/audio retrieval, respectively. Furthermore, we note that GRACE \cite{dis} enables generative text-to-image retrieval, so we add it to the baseline for generative cross-modal retrieval.

\section{Implementation Details}
\label{sec:sig}

\paragraph{Semantic Identiﬁer Generation Pipeline} 
We use ImageBind to unify the processing of multimodal data into a latent space, ultimately obtaining 1024-dimensional embeddings. We employ the default mini batch K-Means algorithm in scikit-learn \cite{scikit}, where $K$ = 128, to cluster 1024-dimensional item embeddings. The encoder in RQ-VAE has three intermediate layers of size 512, 256 and 128 with ELU activation, and final latent representation dimension of 64, with which the deocoder is completely symmetric. There are 2 levels of codebook in the quantizer, and for each level, a codebook with cardinality 128 is maintained, where each vector in the codebook has a dimension of 64. Because we use a prefix database to ensure that each identifier is unique, this means at least $128^3 = 2,097,152$ items can be represented. We train the RQ-VAE using the Adam optimizer with an initial learning rate of 1e-6 and Inverse Square Root scheduler integrated in fairseq \cite{fairseq}. The learning rate increases to 1e-4 after 300 warm-up epochs, and then gradually decreases until 500 epochs. 

\paragraph{Retrieval Pipeline} We initialize all parameters with Xavier uniform distribution. We use Adam optimizer with an initial learning rate of 1e-6, while increasing the learning rate to 1e-4 in 5 epochs using the Cosine Annealing Warmup Restarts scheduler. We set the scaling factor of the consistency-based regularization loss as $\omega$ = 0.15.

\section{Metrics}
\label{sec:metrics}
Recall$@K$ measures how often the desired item is hit in the top $K$ retrieved candidates, where we set $K$ to $1$, $5$ and $10$ in experiments. MRR calculates the reciprocal of the rank at which the first relevant item is retrieved. A high MRR indicates that the relevant item has a high ranking position.

\section{Ablation Study}
\label{sec:ablation}

\begin{table*}[ht]
  \caption{Ablation Study on Text-to-Audio/Video retrieval task.}
  \label{tab:more_ablation}
  \resizebox{\textwidth}{!}{
  \begin{tabular}{cccccccccc}
    \toprule
    \multirow{2}{*}{Setting} & \multicolumn{4}{c}{Clotho} && \multicolumn{4}{c}{AudioCaps} \\ \cline{2-5} \cline{7-10} 
    & R@1 & R@5 & R@10 & MRR@10 && R@1 & R@5 & R@10 & MRR@10 \\
    \midrule
    w/o consistency loss & 44.51 & 68.76 & 74.56 & 49.87 && 47.86 & 76.74 & 83.89 & 57.68 \cr
    w/o fusion strategy & 43.44 & 66.15 & 73.03 & 48.92 && 46.04 & 76.96 & 82.21 & 55.01 \cr
    w/o K-Means & 45.81 & 69.31 & 74.05 & 51.59 && 47.21 & 77.11 & 85.19 & 52.88 \cr
    w/o RQ-VAE & 44.39 & 68.91 & 75.11 & 49.51 && 44.94 & 73.27 & 82.45 & 50.34\cr
    CART & \textbf{46.4} & \textbf{70.6} & \textbf{76.0} & \textbf{52.3} && \textbf{49.8} & \textbf{79.8} & \textbf{86.8} & \textbf{62.0} \\
    \midrule
    \multirow{2}{*}{Setting} & \multicolumn{4}{c}{MSR-VTT} && \multicolumn{4}{c}{MSVD} \\ \cline{2-5} \cline{7-10}  
    & R@1 & R@5 & R@10 & MRR@10 && R@1 & R@5 & R@10 & MRR@10 \\
    \midrule 
    w/o consistency loss & 51.5 & 75.1 & 82.7 & 61.53 && 61.19 & 87.16 & 92.38 & 72.03 \cr
    w/o fusion strategy & 48.4 & 70.3 & 80.5 & 56.38 && 60.75 & 86.57 & 91.64 & 71.71 \cr
    w/o K-Means & 50.1 & 73.1 & 80.9 & 59.79 && 61.04 & 85.52 & 91.34 & 71.69\cr
    w/o RQ-VAE & 46.5 & 70.1 & 78.9 & 56.62 && 58.95 & 82.98 & 87.91 & 68.65\cr
    CART & \textbf{52.6} & \textbf{75.4} & \textbf{84.2} & \textbf{61.77} && \textbf{63.58} & \textbf{87.91} & \textbf{92.84} & \textbf{73.59} \\
    \bottomrule
  \end{tabular}}
\end{table*}

As shown in Table~\ref{tab:more_ablation} and ~\ref{tab:ablation_coco}, we demonstrate the effectiveness of each module in CART on the text-to-image/audio/video retrieval task.

When \textbf{w/o consistency}, the model undergoes overfitting which also leads to degradation of retrieval performance.

When \textbf{w/o fusion strategy}, the metrics show different degrees of decrease, proving the effectiveness of the coarse-to-fine feature fusion mechanism.

When \textbf{w/o K-Means token} or \textbf{w/o RQ-VAE token}, the retrieval performance also decreases, which proves the effectiveness of coarse-fine token.

\begin{table}[t]
\centering
  \caption{Ablation Study on MS-COCO.}
  \label{tab:ablation_coco}
  \resizebox{\columnwidth}{!}{
  \begin{tabular}{ccccc}
    \toprule
    Setting & R@1 & R@5 & R@10 & MRR@10 \\
    \midrule
    w/o consistency loss & 49.86 & 75.69 & 83.73 & 60.02 \\
    w/o fusion strategy & 43.41 & 69.27 & 79.12 & 54.34 \\
    w/o K-Means & 48.11 & 74.87 & 84.08 & 59.53 \\
    w/o RQ-VAE & 51.03 & 76.94 & 84.97 & 62.12 \\
    CART & \textbf{52.42} & \textbf{77.53} & \textbf{86.11} & \textbf{63.19} \\
    \bottomrule
  \end{tabular}}
\end{table}

\section{Decoder Layers}
\label{sec:decoder}
In the previous section, we perform ablation experiments on the number of encoder layers, and correspondingly, we perform ablation experiments on the number of decoder layers. 

As shown in Table~\ref{tab:decoder_layer}, as the number of decoder layers increases, the capability of the model gradually increases and the retrieval performance gradually improves, but too many decoder layers can lead to the model undergoing serious overfitting, leading to performance degradation.

\begin{table}[h]
  \caption{Results with different number of decoder layers.}
  \begin{tabular}{c|c|c|c|c}
    \toprule
    Layers & R@1 & R@5 & R@10 & MRR@10 \\
    \midrule
    1 &  79.48 & 94.66 & 97.02 & 85.96  \\
    2 &  \textbf{82.52} & 95.86 & 98.04 & \textbf{88.31} \\
    3 &  81.78 & \textbf{96.14} & \textbf{98.38} & 88.04 \\
    4 &  73.18 & 92.66 & 96.61 & 81.62\\
    5 &  67.32 & 92.22 & 96.54 & 77.86 \\
    \bottomrule
  \end{tabular}
  \label{tab:decoder_layer}
\end{table}

\section{Identifier Configuration}
\label{sec:idconfig}

As shown in Table~\ref{tab:size} We try to vary the $K$ in K-Means as well as the codebook size in RQ-VAE, and observe that the retrieval performance of the model is more robust for most settings. Only when $K$ is set to 64 or codebook size is set to 64, the model performance shows a more significant degradation. We analyze that too small a cluster category will lead to lower differentiation of K-Means and fail to achieve good clustering effect, and too small codebook size will likewise lead to limited semantic differentiation ability, resulting in larger information loss. Furthermore, it will also lead to an increase in the amount of items with the same identifiers, thus the model can only rely on the unique token which do not have any semantic information. We try to add codebook in RQ-VAE to increase the length of semantic identifiers and we observe that the retrieval performance fluctuates only slightly, as shown in Table~\ref{tab:length}. This demonstrates the effectiveness of the semantic identifiers and the stability of the retrieval model. Considering that too long semantic identifiers lead to lower codebook utilization and increased decoding steps, we only use four tokens. It is worth noting that retrieval performance exceeds the baseline in all settings, despite varying the length and size of the identifiers.

\section{Codebook Dimension}
\label{sec:dim}
We further explored the codebook dimension by keeping the number of clustering centers at 128 and conducting experiments on Flickr30k, as shown in Table~\ref{tab:dimensionality}. We note that large dimensions not only make the samples sparse and affect the distance discrimination, but also may introduce more noise and affect the model learning semantics. And small dimensions will reduce the learning ability of codebook and make it difficult to revert the data features. Note that these parameter settings are beneficial for this experiment, and there may be some mathematical connection between them, which future work could explore more to help determine the best parameters for different datasets.

\begin{table}[h]
  \caption{Results for different settings of codebook dimension.}
  \centering
  \resizebox{\columnwidth}{!}{
  \label{tab:dimensionality}
  \begin{tabular}{c|c|c|c|c}
    \toprule
    Dimensions & R@1 & R@5 & R@10 & MRR@10 \\
    \midrule
    32 & 77.84 & 93.62 & 96.58 & 84.66 \\
    64 & 81.78 & 96.14 & 98.38 & 88.04\\
    128 & 79.88 & 94.52 & 97.02 & 85.91  \\
    \bottomrule
  \end{tabular}}
\end{table}

\end{document}